\documentclass[fleqn,twoside]{article}
\usepackage{espcrc2}

\usepackage{graphicx}

\newcommand{\be}{\begin{equation}}
\newcommand{\ee}{\end{equation}}

\newcommand{\AmS}{{\protect\the\textfont2
  A\kern-.1667em\lower.5ex\hbox{M}\kern-.125emS}}

\title{Hadronic transitions from the lattice}

\author{UKQCD Collaboration: C. Michael\address{Theoretical Physics Division, 
Dept. of Mathematical Sciences,  \\       
  University of Liverpool, Liverpool L69 3BX, UK }
 }

\begin{document}

\begin{abstract}
 I discuss strategies to determine hadronic decay couplings from 
lattice studies. 
 As a check of the methods, I explore the decay of a vector meson to two
pseudoscalar mesons with $N_f=2$ flavours of sea quark.
Although we are working with quark masses that do not allow a physical
decay, I show how  the transition rate can be evaluated from the 
amplitude for $\rho \to \pi \pi$  and from the annihilation component of
$\pi \pi \to \pi \pi$. I explore  the decay amplitude for two different
pion momenta and find consistent results. The coupling strength  found
is in agreement with experiment.  I also find evidence for a shift in
the $\rho$ mass caused  by mixing with two pion states. 
 I also present results for the decay of a hybrid meson, for the case 
of heavy valence quarks.
\end{abstract}

\maketitle

\section{INTRODUCTION}


Rather few hadronic states are actually stable under strong interactions. 
Because of this, a full understanding of hadrons from QCD  will involve 
developing techniques to treat unstable hadrons and to evaluate their 
hadronic transition strengths. 

An area of hadronic physics that is of considerable interest is that  of
gluonic excitations: the exploration of hadrons with a non-trivial
gluonic content.  The most clear case is that of glueballs, however, any
glueball states will mix  with quark-antiquark mesons in practice. This
mixing can be thought of as mediated by a hadronic transition. Moreover,
the eventual hadronic decays of such mixed mesons also need to be
evaluated  to give a clearer picture of the relationship between the
physical spectrum  and the ingredients (glueballs and quark-antiquark
mesons). 

 The other natural area to explore gluonic excitations is that  of
hybrid mesons. These are quark-antiquark mesons with an additional 
non-trivial gluonic component. The benchmark example is a spin-exotic
meson  (so-called because it cannot be made from combining a quark and
antiquark). These  spin-exotic hybrid mesons will also be unstable, and
study of their decay  transitions will guide the interpretation of
experimental candidates.

 An understanding of methods to treat  purely hadronic transitions will
also help in developing methods to  explore weak decays to hadronic
final states (e.g. $K \to \pi \pi$).

The lattice QCD approach is the only quantitative non-perturbative
method  but the formulation in  euclidian time causes problems with the 
treatment of decays as I now discuss.

 \subsection{Euclidian time}

Hadron masses are extracted from lattice QCD calculations using two
point correlators, however,  their behaviour as $e^{-Mt}$  in euclidian 
time  will  not be appropriate for hadrons that can decay. The most
naive modification of the lattice QCD formalism caused by the
introduction of decay widths is the replacement  $M \rightarrow M -
i \Gamma /2$ where $\Gamma$ is the decay width.
 This gives a behaviour as 
$  e^{ -M t } e^{ i\Gamma  t/2 }    $
 but this is not consistent: the correlation must be  positive
definite yet this expression oscillates. Here I discuss this apparent
paradox  more fully~\cite{cmdecay}.

 The euclidian time propagator of an unstable particle which can decay
into two stable particles of  mass $\mu$ can be expressed as
 \be
G(t) =  \frac{1}{\pi}  \int_{2 \mu}^{\infty} dE e^{-E t} \rho(E)
 \label{eq.euclid}
\ee
 where we can express the spectral density for a resonance of mass $M$ as
 \begin{equation}
\rho(E)= \frac{1}{2 M} \  \frac{ \Gamma(E)/2  }{ (M-E)^2 + (\Gamma(E)/2)^2  }
 \end{equation}

As an illustration, consider taking $\Gamma(E)$ as a constant with
 $\Gamma  \ll  (M - 2 \mu)$ and  $\ ( M - 2 \mu ) t \gg 1 $. Then we have,
in this approximation, 
 \begin{equation}
 2M G(t) = e^{ -M t } \cos(\Gamma t/2 ) + 
 \frac{ e^{- 2 \mu t } \Gamma} { 2 \pi  (M-2\mu)^2  t }
 \label{cmn:eq:corrRESmodel}
 \end{equation}
 This expression, when evaluated, shows that the threshold contribution
(the second term)  dominates at larger $t$ and hence the oscillating
behaviour of the  first term will be obliterated. This clarifies the
apparent paradox mentioned above concerning the sign of $G(t)$: the 
contribution from the lower lying (in energy) component of the resonance
is  enhanced in euclidian time. Indeed this can be seen directly from
eq.~\ref{eq.euclid}.

 By contrast, the Minkowski behaviour of the two particle correlator
under the  same assumptions is 
 \begin{equation}
 2M G_M(t) = 
e^{ -iM t } e^{-\Gamma t/2 } - 
 \frac{i e^{ -2i \mu t } \Gamma} { 2 \pi  (M-2\mu)^2  t }
 \end{equation}
 for which the second term is almost always negligible.

 In principle, a lattice determination of the spectral function
$\rho(E)$  from measurements of $G(t)$ using eq.~\ref{eq.euclid} would be
sufficient to give a full description of hadronic decays. This is  not a
simple task: the inverse Laplace transform need to obtain $\rho(E)$ from
$G(t)$ is not numerically stable. It has to be stabilised by  making
model assumptions about $\rho(E)$, for example as discussed in
refs.~\cite{cmdecay,mart}. 
 Another possibility is that the maximum entropy method 
may provide sufficient numerical stability and this has been tested 
in model cases~\cite{Yamazaki:2002ir}.

 The same observation that the region near to threshold  dominates is at
the root of the conclusions~\cite{mt} of Maiani and Testa that two body
states  have unappetizing properties in euclidian time.

\subsection{Particles in a box}

 In practice this issue is less relevant since the lattice evaluations
are performed in  a finite spatial volume with periodic boundary
conditions.  This finite spatial size of the lattice implies that
two-body  states are actually discrete. By measuring their energy very
precisely as the spatial  volume is varied, it is
possible~\cite{luscher} to extract the scattering phase shifts and hence
decay properties.

 Consider the two body states of two pions of mass $\mu$ with  momentum
$\pm {\bf k}$ which will have total energy $E=2(\mu^2+{\bf k}^2)^{1/2}$.
 Then if they are non-interacting, their momentum on a spatial
hyper-torus of size $L$ will be discrete with ${\bf k}=2\pi {\bf n}/L$
where ${\bf n}$  has integer components (and for future use we define
$N$ such that there  are $N$ values of ${\bf n}$  having a given
$n=|{\bf n}|$, so $N=6$ if $n=1$). 
 If the two pions interact with a finite range of interaction, compared 
to the spatial volume, this implies that the finite size effect will result 
in an momentum shift of order $L^{-3}$, provided that $L$ is large enough
to contain one  pion without undue distortion.
The detailed analysis of L\"uscher~\cite{luscher} expresses the shift 
in this momentum value when there are interactions among the pions with 
phase-shift $\delta(k)$, provided $E$ is below the inelastic threshold, as
 \be
  d(E^2) = 4 d(k^2) = {-16 \pi N \tan \delta(k) \over L^3 k}
 \ee
 to leading order in $L^{-1}$, and indeed a much more precise expression
 with sub-leading terms  is available through L\"uscher's
work~\cite{luscher}.

 Now when there is a resonance at $m_R$ near to a two particle state,
there will be  an opportunity to deduce the resonance width. Then, one
can describe the phase shift  $\delta(k)$ by using the expression for
elastic $\pi \pi$ scattering dominated by  this resonance pole: 
 \be
 \tan \delta(k) = { \Gamma(k) \over 2 (m_R-E)}
 \label{eq.pole}
 \ee
 so that determinations of $\delta(k)$ at two or more values of $k$ will 
enable the resonance parameters $m_R$ and $\Gamma(k_R)$ to be evaluated, 
where $m_R^2=4(\mu^2+k_R^2)$.  
  The most promising way to determine the two particle energy accurately
 is by using several lattice operators (including some built like  two
pions and some built like the resonance under study)  as sources and
sinks to get the best determination of the energy levels from fitting
the matrix of correlators.  Note that in practice it will be difficult
to  obtain sufficiently accurate energy determinations of the lowest
two-particle  state and it will be even more difficult to determine the
energy level  of the next heavier state with larger momenta.

\subsection{A direct approach}


 It should be possible to extract the hadronic transition amplitude 
directly  from the lattice (rather than via energy level shifts), and
this we now discuss.

 Using suitable lattice operators to  create $R$ at $t=0$ and annihilate
a two-pion state with momenta $k$ and $-k$ at  time $t$,  the 
contribution to the correlator from a $R$ state with mass $m_R$  and a
two-body state with energy $E_{\pi \pi }$ is given to leading order in
$x$  by
 \be
   C_{R-\pi \pi } (t)=\sum_{t_1} h e^{-m_R t_1} x e^{ -E_{\pi \pi } (t-t_1)} b
 \label{3pt}
 \ee
 where there is a summation over the intermediate $t$-value  $t_1$  and 
where $h$ and $b$ are the amplitudes to make each state from the lattice
operators used  and $x$ is the required transition amplitude $\langle R
| \pi \pi  \rangle$.  Here we are assuming that the states $R$ and $\pi
\pi $  are normalised to unity.  By obtaining $h$ and $b$  from the $R
\to R$ and $\pi \pi  \to \pi \pi $ correlators,  one can attempt to 
extract $x$.

 The complication, however, is that removal of excited state
contributions is  tricky, even if variational methods are used  to
construct improved lattice operators to create the ground states.   For
example, if $m_R - E_{\pi \pi }  > 0 $ then the transition time $t_1$ 
will be preferentially near 0 (since the heavier state then propagates
less  far in time) and one can complete the sum over $t_1$ obtaining a 
$t$-dependence of eq.~\ref{3pt} as $e^{-E_{\pi \pi }t}$. This same
$t$-dependence  would be obtained if the state with mass $m_R$ were to
be replaced with an  excited state with an even heavier mass. Thus one
cannot separate the  ground state and excited state contributions even
in principle. See ref.~\cite{cmcm,hdecay} for  a fuller discussion.

 The way forward is that if $m_R = E_{\pi \pi }$, the ground state
contributions to eq.~\ref{3pt} have a $t$-dependence  as $t e^{-E_{\pi
\pi }t}$ whereas any excited state contributions behave as $e^{-E_{\pi
\pi }t}$  as above. So now we do have a way to isolate the required
ground state  contribution:
 \be
    xt = 
   {C_{R-\pi \pi } (t) \over  [C_{R-R} (t)\ C_{\pi \pi -\pi
\pi } (t)]^{1/2} } + {\cal O}(x) + {\cal O}(x^3 t^3)
 \label{xrat}
 \ee

Note that this separation is only by a power of $t$ which is  less than
the case for two-point  correlations where the excited state 
contributions are suppressed by an exponential $e^{-(m'_R-m_R)t}$.  
 
  In practice the requirement of energy equality can be relaxed.
Defining $\Delta=  m_R - E_{\pi \pi }$, then the ground state contribution
to the expression of eq.~\ref{xrat} evaluates  to   $2 x \sinh
(t\Delta  /2)/(t\Delta ) = x(1+(t\Delta )^2/24 + \dots)$. So this will be
equivalent to the expression  with $\Delta=0$ provided
 \be
   (m_R - E_{\pi \pi }) t \ll 5
 \label{onshell}
 \ee

\begin{figure}[h]

\begin{center}
  \includegraphics[scale=0.3]{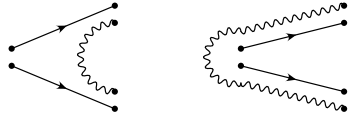}

 \caption{ Quark pair production (wiggly lines) for the three point
function $R \to \pi + \pi$  in euclidian time (running horizontally). The
left hand diagram has  the interpretation of a transition (our $x$) at
an intermediate  time while the right hand diagram can be thought of as 
some intrinsic mixing in the $R$ state.
 }
 \label{3ptfig}

 \end{center}
\end{figure}

 So far we have described the behaviour of the  $C_{R-\pi \pi } (t)$ in
the limit of small $x$. The first correction term from multiple
transitions (to eq.~\ref{xrat}) will be of order  $x^3t^3/6$ so
we need $xt \ll 1$.  As well as these  corrections  which  are of higher
order in $xt$, one must also consider the intrinsic mixing of the
initial $R$ state with $\pi \pi $ (and vice versa). This intrinsic
mixing (i.e. coming from the  lattice operators used to create the
states having an admixture - see  fig.~\ref{3ptfig} for an example) is 
expected be of order $x/{\cal E}$ where ${\cal E}$ is the energy of the
quark pair and so  will contribute a term like $x e^{-E t}/{\cal E}$.
This is a contribution  similar to that from excited states  and so will
be dominated at large $t$ by the $xt e^{-E t}$ term we are looking for.
So we need both $xt$ to be small and $t$ to be large. This implies that 
$x$ must be small for this direct approach. And indeed, it is only when
$x$ is small that the more  rigorous approach of determining the energy
shifts will be computationally  difficult to implement.

 In summary, when  $t (m_R - E_{\pi \pi } ) \ll 5$;  $xt \ll 1$ and
$(E'-E)t \gg 1$ (with $E'-E$ the energy gap to  the first excited
state),  it is possible to estimate the hadronic transition amplitude
directly, and this  we will explore in detail for $\rho$ decay.

\subsection{Comparison}

 In order to show the relationship of this direct approach with
L\"uscher's approach, consider  the case when  a two particle state is
close in energy  to the resonance.  It should be possible to arrange
that energy levels are sufficiently  close by choosing the lattice
spatial length $L$.
 Then once $x$ has  been  determined in this situation where the
resonance does not decay significantly (because  the two particle level
has the same energy as the resonance), one can assume that  the coupling
strength determined can be used in a larger spatial volume  where decay
is energetically allowed.

 The most direct way to evaluate resonance decay is using  first order
perturbation theory (Fermi's Golden Rule) which implies  a transition
rate 
 \be 
 \Gamma=2 \pi \langle x^2 \rangle \rho(E)
 \label{fgr}
 \ee
  where the angle brackets indicate that an average over spatial
directions will be needed. For a decay from the centre of mass with
relative momentum $k$, the density of states  $ \rho(E)= L^3 k E /(8
\pi^2)$.

 Now to compare with L\"uscher's approach, we need to evaluate the
energy shift  of the two particle state in a finite box (which is
exactly $N-$fold degenerate).  This can be estimated from the mixing of
the  two energy levels  which are close to each other, using second
order  perturbation theory (in $x$) which gives an energy shift, on the
lattice, of
 \be
   dE = - {N x^2 \over m_R-E},
 \ee
  Thus when the resonance lies  above the two particle state, the mixing
will move the resonance mass up  and the two particle energy down, in
the usual way as mixing causes repulsion of levels.

 Using the relationship between $x$ and $\Gamma$ from eq.~\ref{fgr},  we
obtain 
 \be
   dE = -{4 \pi N \Gamma \over (m_R-E) k E L^3}  
 \ee
  which is exactly the same expression as would be obtained at leading
order in  $L^{-1}$ from L\"uscher's formalism as presented above. This
is as it should be, but does  illustrate that an estimate of $x$
directly from the lattice may  provide a useful way to evaluate these
decay effects.

 Indeed the mass shifts are of second order in $x$ so will be small in
general,  and hence difficult to determine with sufficient precision.
The direct measurement   of the transition giving $x$ from the  lattice is
thus an attractive prospect.

\section{VECTOR MESON DECAY}

\subsection{Transitions}

For on-shell transitions,  it is  possible to estimate hadronic
transition strengths  as described above and here we explore this
approach further  for the case of  $\rho$ meson decay to $\pi \pi$,
following ref.~\cite{rhodecay}.

\begin{figure}[th]
 \vspace{-3cm}  
  \includegraphics[scale=0.3]{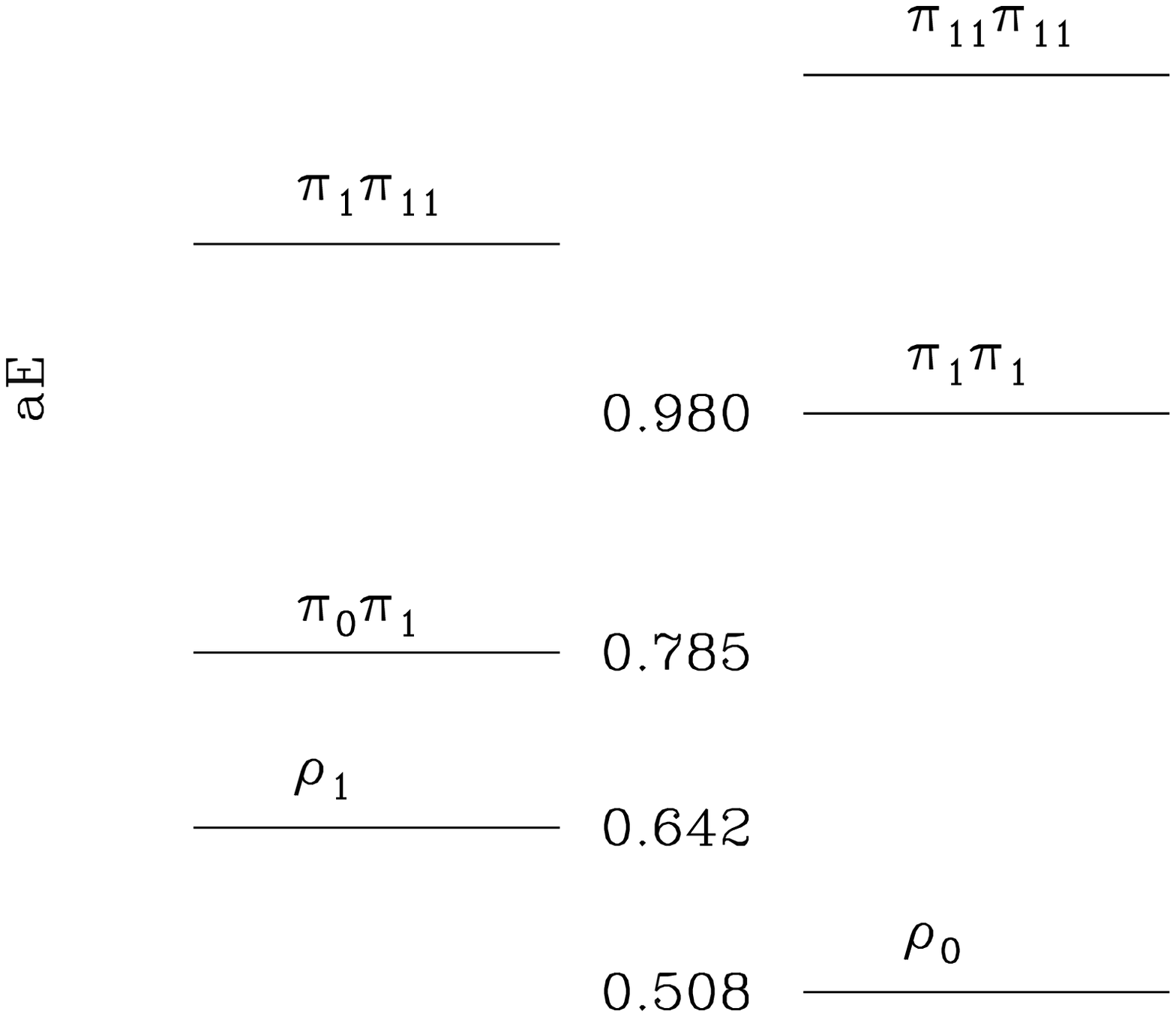}
 \vspace{-2.5cm}
  \caption{}
 \label{fig.rpp}
\end{figure}

 The situation we shall analyse is represented by the energy spectrum 
shown in fig.~\ref{fig.rpp}, here neglecting any interactions among the
states. Note that no decay can actually take place with these lattice
parameters. This is , however, optimum for our analysis since when the 
energy levels are approximately degenerate, the hadronic transition
amplitude can be  evaluated more effectively.

\begin{figure}[th]
  \includegraphics[scale=0.4]{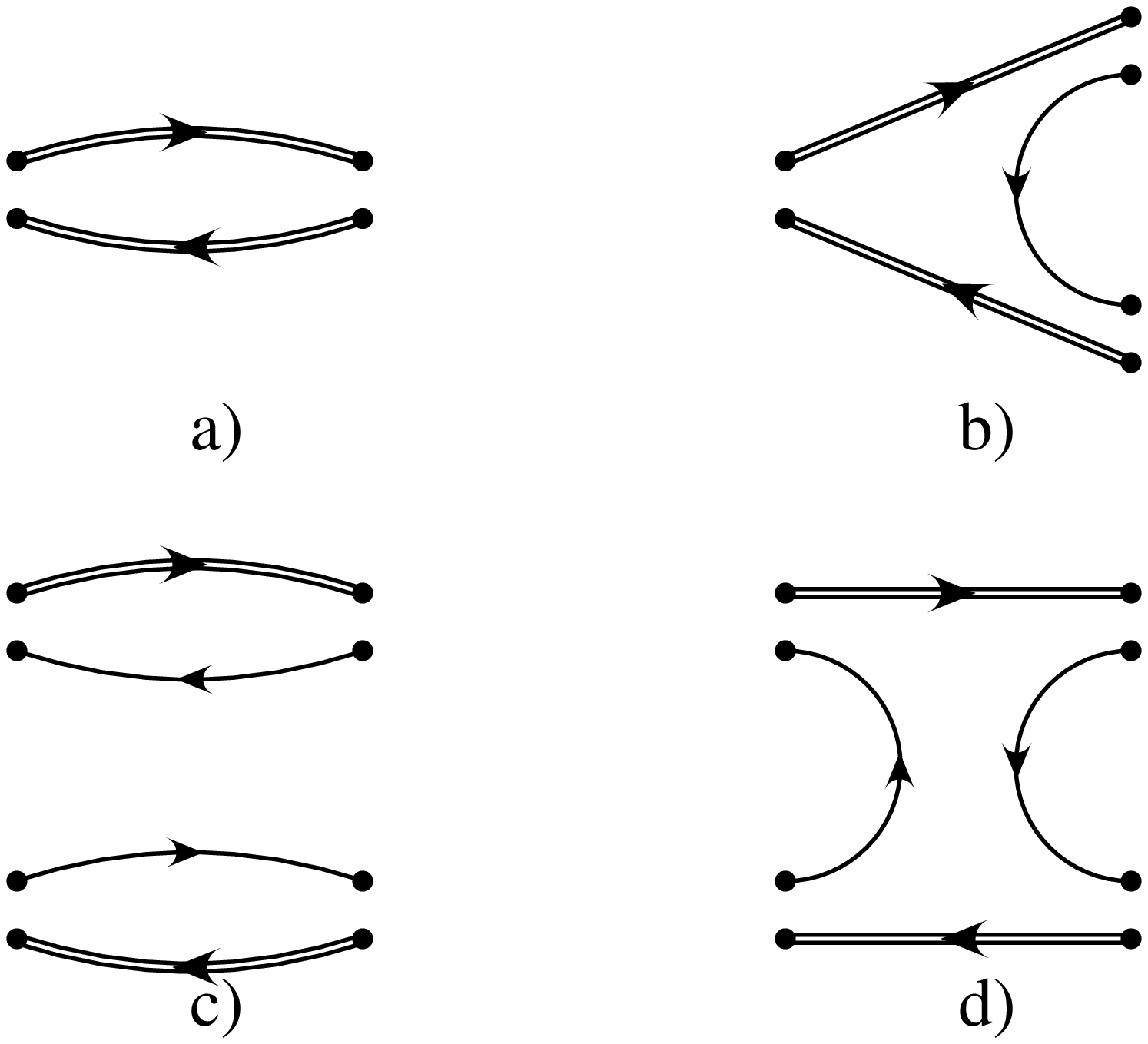}
\vspace{-1.0cm} 
 \caption{  }
 \label{fig.quark}
\end{figure}

\begin{figure}[th]
 \vspace{-2.0cm}
  \includegraphics[scale=0.35]{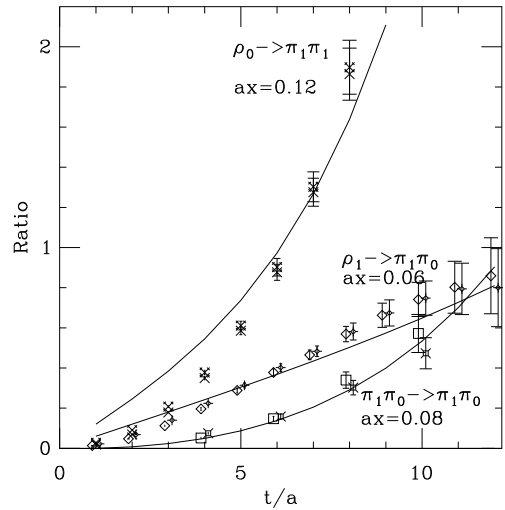}
 \vspace{-1.5cm} 
 \caption{Ratios of 3 and 4-point correlators.}
 \label{fig.xtall} 
 \end{figure}

We evaluate correlations between lattice operators creating both
a $\rho$ meson and a $\pi \pi$ state, using a stochastic  method with
sources on time-slices~\cite{fm} to evaluate the quark diagrams shown in
fig.~\ref{fig.quark} from 20 gauge configurations~\cite{ukqcd} with
$N_f=2$ sea quarks  of mass corresponding to about 2/3 of the
strange-quark mass. 
  These correlations, normalised by the two point functions as
appropriate,  are illustrated in fig.~\ref{fig.xtall}.  Here the
off-diagonal case  (labelled $\rho_1 \to \pi_1 \pi_0$) shows the
important feature that it grows  approximately linearly with increasing
$t$. As emphasized above, this  linear growth will only
occur for on-shell transitions and this is  essentially the case here.

 To extract an estimate of the transition amplitude, we wish to 
evaluate $x=\langle \rho | \pi \pi \rangle$ where these states are 
normalised on the lattice (to unity).  If higher excited states are
neglected,  one can make a two state model (i.e. $\rho$ and lightest
$\pi \pi$ state)  with this transition amplitude and evaluate the
contributions as shown by the curves in fig.~\ref{fig.xtall}. Because
the location (at $t_1$ between 0 and $t$) of any transition is not
known, one has potentially very  serious contributions from excited
states. As  discussed above, these excited state contributions can  be 
avoided if the transition is approximately on-shell, when the  linear
dependence on $t$ of the off-diagonal transition (with value $xt$) is a
unique  signature of this on-shell transition that we wish to extract.

 Note that for the case  of $\rho_0 \to \pi_1 \pi_1$, the on-shell
condition is much less well satisfied but the relative momentum of the
pions in the centre of mass is twice as large as for $\rho_1 \to \pi_0
\pi_1$ and hence  $x$ should be approximately twice as large (as we
find), since for a P-wave decay there will be a momentum factor  in the
transition amplitude.

 A further check of the extraction of $x$ comes from the  box diagram, 
fig.~\ref{fig.quark}d,  which will have a contribution  behaving as $x^2
t^2/2$ arising from a $\rho$ intermediate state - see
the data labelled $\pi_1 \pi_0 \to \pi_1 \pi_0$ in fig.~\ref{fig.xtall}.
Again this contribution does seem to be present at the required level.

Thus we have qualitative agreement from the three-point evaluations with 
two different relative momenta and from the four-point evaluation that 
$xa \approx 0.06$.

 A more quantitative estimate of $x$ can be made~\cite{rhodecay} by
reducing excited state contributions as much as possible.  Thus one uses
 fits to remove excited state contributions from the  normalising
two-point functions and  then evaluates the slope of the ratio of
eq.~\ref{xrat} versus $t$ to  remove the ${\cal O}(x)$ term also.

 Note that the above methods can be used in the quenched
approximation~\cite{wein} to estimate hadronic transition amplitudes. 
They depend on assuming that $xa$ is fairly weak, as we indeed find.

\subsection{Energy shifts}

 A more rigorous approach is to focus on energy values. When two levels
are close (our on-shell condition), then they will mix and the resultant
energy  shifts give relevant information~\cite{luscher} as discussed
above.  Moreover we can estimate these  energy shifts from our $x$-value
which provides more cross checks. These energy shifts can only be
studied using dynamical fermions.

\begin{figure}[th]
 \vspace{-2.0cm}
  \includegraphics[scale=0.35]{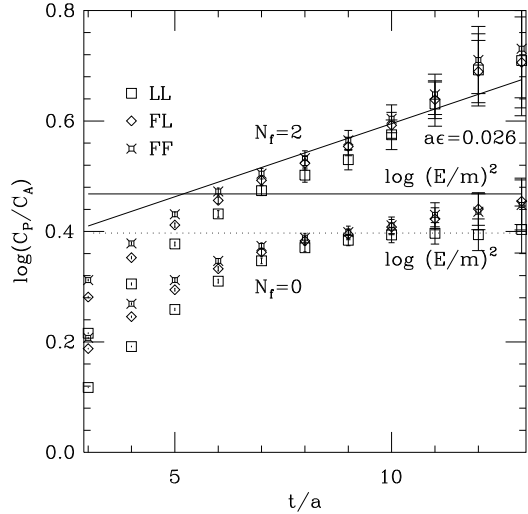}
 \vspace{-1.5cm}
 \caption{The ratio of Parallel to Antiparallel $\rho$ correlators.} 
 \label{fig.rhoratall}

\end{figure}

From a full  analysis (i.e. measuring all the quark diagrams illustrated
 in fig.~\ref{fig.quark} and fitting the matrix of correlators) we
obtain the energy shift of the $\pi_1 \pi_0$ state (i.e. the un-binding
energy), as needed in L\"uscher's approach, as 0.02(2) (in lattice
units) upward which is consistent but not sufficiently accurate to be of
significant use. The energy shift of the $\rho_1$ state can, however, be
determined in this case because of a lattice artifact. The $\rho$ with
momentum 1 (in lattice units of $2 \pi/L$) can have its spin aligned
parallel to the momentum axis (P) or perpendicular to it (A). Because
the $\pi_0 \pi_1$ state has relative momentum along a lattice axis and
the transition from $\rho$ to $\pi \pi$ has orbital angular momentum L=1
(so a distribution like $\cos \theta$ in the centre of mass), only the
parallel state (P) can mix with this two pion state. This  situation is
not typical - it occurs because the $\rho$ with momentum 1  only has a
transition to a  $\pi_0 \pi_1$ state with a pion having momentum in the
same direction.
  This mixing will not be present in the quenched approximation, so this
provides a direct opportunity to see the effect of the two pion channel
on the $\rho$ in unquenched studies. We do indeed find~\cite{rhodecay}
such a mass splitting between the P and A orientations of the $\rho$ 
due to mixing for $N_f=2$ but not for $N_f=0$ as shown in
fig.~\ref{fig.rhoratall}. Moreover the magnitude of this energy shift
($0.026\pm 7$ in lattice units) is consistent with other determinations
of the transition strength.

\subsection{Phenomenology}

 The basic assumption is that the transition from $\rho$ to $\pi \pi$ 
is given by an effective interaction with a finite spatial extent, this
is  usually summarised by an effective lagrangian where we normalise the
coupling  as $\bar{g}^2=\Gamma M E/k^3$ in terms of the decay width.
 Then, provided the lattice spatial size is big enough that the hadrons
are not distorted,  our lattice situation (where no decay occurs) can be
used  to determine $\bar{g}$ and this can then be used to predict decay
widths  when the lattice volume is increased so that the minimum
momentum  becomes small. In our case we will also need to reduce the 
quark mass to allow decay since we have $m_{\pi}/m_{\rho} =0.58$.

 One complication is that the transition which is closest to being 
on-shell is $ \rho_1 \to \pi_1 \pi_0$ which is not in the centre of
mass.  A generalisation of L\"uscher's approach has been
made~\cite{rumm} which allows  for this. The more direct method we have
described here can  also be extended to this case~\cite{rhodecay}. There
are possible problems in this treatment of decays of moving  particles.
For instance, on the lattice a boost to bring the $\rho$ to rest will 
not bring the two pion state to have zero net momentum also. This is 
because the lattice situation allows energy non-conservation which, 
upon  boosting, will imply some momentum non-conservation.  
 
 From our lattice studies, we deduce that $ax=0.06^{+2}_{-1}$ for 
$\rho_1 \to \pi_1 \pi_0$. Translating~\cite{rhodecay} this lattice
transition amplitude to the  continuum normalisation, gives
$\bar{g}=1.40^{+27}_{-23}$. Using the observed $\rho_1$ energy shift 
gives another estimate, namely $\bar{g}=1.56^{+21}_{-13}$.   These two
values agree well, which is good support for our scheme of extracting
hadronic  transition strengths.

Note that our lattice values  would need to be extrapolated to light
sea-quarks and to the continuum limit  to allow all sources of
systematic error to be explored in the comparison with the experimental 
situation. Nevertheless, these  two values agree well with the values
extracted from experimental data on decays of $\rho$, $K^*$ and $\phi$
mesons, namely $\bar{g} \approx 1.5$. This further underlies the 
viability of our methods.

\section{HYBRID MESON DECAY}

 Hybrid mesons are those with non-trivial gluonic excitations. The spectrum 
of such mesons has been determined from lattice studies, both for the case 
of heavy quarks and light quarks. 
 Hybrid mesons have allowed couplings to meson meson systems. If these 
coupling are large, then the hybrid mesons decay widths  will be big and
such mesons will be difficult to identify  experimentally.
 Here we present recent lattice results on these  hadronic transitions.

\subsection{Hybrid mesons with heavy quarks}

 One well defined way to treat heavy quarks (such as the $b$-quark) on a
 lattice is to use the leading order of the HQET which corresponds to 
the static approximation. Here we take  the heavy quarks to be fixed at
locations  $R$ apart in the $z$-direction.

Consider $Q \bar{Q}$ states with static quarks  in which the gluonic
contribution may be excited. We  classify the gluonic fields according
to the symmetries of the system.  This discussion is very similar to the
description of electron wave functions in  diatomic molecules. The
symmetries are  (i) rotation around the separation axis $z$ with
representations labelled by $J_z$ (ii) CP with representations labelled
by $g(+)$ and $u(-)$ and (iii) C${\Re}$. Here  C interchanges $Q$ and
$\bar{Q}$, P is parity and ${\Re}$ is a rotation  of $180^0$ about the
mid-point around the $y$ axis. The C${\Re}$ operation is only relevant
 to classify states with $J_z=0$. The convention is to label states of
$J_z=0,1,2$ by $ \Sigma, \Pi, \Delta$  respectively. The ground state
($\Sigma^+_g$) will have $J_z=0$ and $CP=+$.

 The exploration of the energy levels  of other representations has a
long history in lattice studies~\cite{liv,pm}. The first excited state
is found  to be the $\Pi_u$.  This can be visualised  as the symmetry of
a string bowed out in the $x$ direction minus the same  deflection in
the $-x$ direction (plus another component of  the two-dimensional
representation with the transverse direction $x$ replaced by $y$),
corresponding to flux  states from a lattice  operator which is the
difference of U-shaped paths from quark to antiquark of the form $\,
\sqcap - \sqcup$.

Recent lattice studies~\cite{jkm}  have used an asymmetric space/time
spacing which enables excited states to be  determined comprehensively.
 These results confirm the finding that 
the $\Pi_u$ excitation is the lowest lying and hence of most relevance 
to spectroscopy.

 From the potential corresponding to these excited gluonic states, one
can  determine the spectrum of hybrid quarkonia using the Schr\"odinger
equation in the Born-Oppenheimer approximation.  This approximation will
be good if the heavy quarks move very little in the  time it takes for
the potential between them to become established. More  quantitatively,
we require that the potential energy of gluonic excitation is much
larger than the typical energy of orbital or radial excitation.  This is
indeed the case~\cite{liv}, especially for $b$ quarks. Another nice
feature of this approach is that the  self energy of the static sources
cancels in the energy difference between this  hybrid state and the
$Q \bar{Q}$ states. Thus the lattice approach gives directly the
excitation energy  of each gluonic excitation.

  The $\Pi_u$ symmetry state corresponds to  excitations of the gluonic
field in quarkonium called magnetic (with $L^{PC}=1^{+-}$) and
pseudo-electric (with $1^{-+}$) in contrast to the usual  P-wave orbital
excitation which has $L^{PC}=1^{--}$. Thus we expect different quantum
number assignments from those of the gluonic ground state. Indeed
combining with the heavy quark spins, we get a degenerate  set of 8
states:

\begin{table}[h]
\begin{center}
\begin{tabular}{|c|c|ccc|} \hline
  $L^{PC}$ &   $J^{PC}$ &$J^{PC}$ &$J^{PC}$ &$J^{PC}$ \\ \hline
$1^{-+}$ & $1^{--}$ & $0^{-+}$ & $1^{-+}$ & $ 2^{-+}$ \\ 
$1^{--}$ & $1^{++}$ & $0^{+-}$ & $1^{+-}$ & $ 2^{+-}$ \\ 
\hline
\end{tabular}
\end{center}
\end{table}

\noindent  Note that of these,  $J^{PC}=  1^{-+},\ 0^{+-}$ and  
$2^{+-}$  are spin-exotic and hence will not mix with $Q\bar{Q}$ states.
They thus form a very attractive goal for experimental searches for
hybrid  mesons.

\subsection{Hybrid decays}

 Within this static quark framework, one can explore the decay
mechanisms.  One special feature is that the symmetries of the quark and
colour fields about the static quarks must be preserved exactly in
decay, hence the light quark-antiquark pair produced must respect these
symmetries.   This has the consequence that the decay from a $\Pi_u$
hybrid state to the open-$b$ mesons ($B \bar{B},\   B^* \bar{B},\  B
\bar{B^*},\ B^* \bar{B^*}$) will be forbidden~\cite{hf8,hdecay} if the 
light quarks in the $B$ and $B^*$ mesons are in an S-wave relative to
the heavy quark (since the final state will have the light quarks in
either a triplet  with the wrong $CP$ or a singlet with the wrong
$J_z$). The decay to $B^{**}$-mesons with light quarks in a P-wave is
allowed by symmetry but not energetically.

\begin{figure}[th]
 \vspace{-2.9cm}
  \includegraphics[scale=0.40]{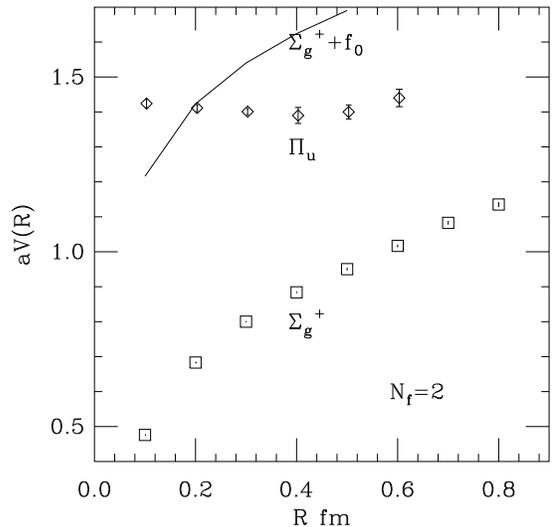}
 \vspace{-1.5cm}
 \caption{The potential energy $V(R)$ (in lattice units with a=0.0972 fm)
versus quark separation $R$ in fm  for  2 flavours of sea quark. The 
energies are given~{\protect\cite{hdecay}}  for the ground state and
first excited gluonic state  and for the two body state of ground state
potential plus scalar meson  ($f_0$) in a P-wave with the minimum
non-zero momentum. The on-shell transition can be evaluated when $R
\approx 0.2$ fm.
 }
 \label{ex.hdecay}
\end{figure}

In the heavy quark limit, the only allowed decays are when the hybrid
state de-excites to a  non-hybrid state with the emission of a light
quark-antiquark pair. Since the  $\Pi_u$ hybrid state has the heavy
quark-antiquark in a triplet P-wave state,  the resulting non-hybrid
state must also  be in a triplet P-wave since the heavy  quarks do not
change their state in the limit of very heavy quarks. Thus the decay for
$b$ quarks will be  to $\chi_b +M$ where $M$ is  a light quark-antiquark
meson in a flavour singlet. This proceeds by a disconnected  light quark
diagram and it would be expected~\cite{vall} that the scalar or
pseudoscalar meson  channels are the most important (i.e. they have the
largest relative OZI-rule  violating contributions).  
 This transition can be estimated on a lattice when the initial and
final energies  are similar. This is the case~\cite{hdecay} for the
$\Pi_u$ de-excitation to ground state  gluonic field plus $f_0$ meson
when the interquark separation is around  0.2 fm which allows a lattice
evaluation of the  hadronic transition strength - see
fig.~\ref{ex.hdecay}.  

 Note that in fig.~\ref{ex.hdecay} for $R=0.1$ fm the $\Pi_u$ state is 
unstable to decay and, since we are using $N_f=2$ flavours of sea-quark,
 this decay is enabled and we might expect to see the  lower energy  (i.e.
$\Sigma_g^+ + f_0$) instead for this  energy level.
 In practice, as we shall find later, the transition strength is quite 
small with $x \approx 0.01$ so the admixture of the lighter
state is very small indeed (it is of order $x^2$) and hence it will only
be significant at  very large $t$-values, much larger than those used here.

 \begin{figure}[ht] 
 \vspace{-2.0cm}
  \includegraphics[scale=0.35]{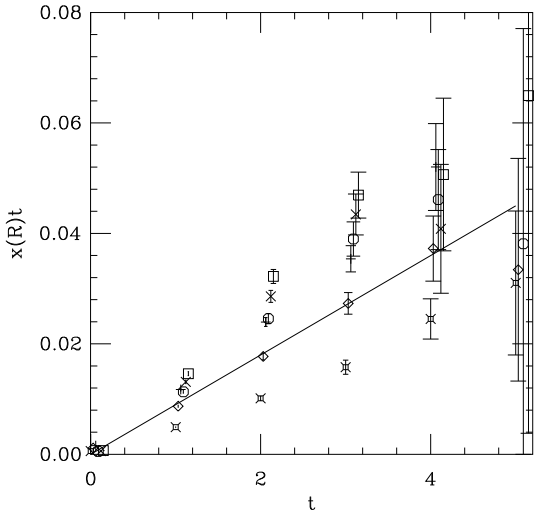}
 \vspace{-1.5cm}
  \caption{ The transition matrix element $xt$  for $\Pi_u \to
\Sigma_g^+ f_0$ with momentum  ${\bf n}=(1,0,0)$ versus $t/a$. Here
$R=0.1$ to 0.6 fm is represented by symbols: fancy square, diamond, +,
octagon, $\times$, square. The line represents a linear  fit to the
$R=0.2$ fm case.
 }
   \label{fig.xhyb}
  \end{figure}

Indeed the dominant transition  is found to  be with $M$ as a scalar
meson and the signal for the  hadronic transition strength $x$ is shown
in fig.~\ref{fig.xhyb}. Making use of the transition strength evaluated
for a range of $R$-values around 0.2 fm and folding in the wavefunction
 overlaps, we can estimate the decay width of the hybrid  spin-exotic
meson $H$ (e.g. $J^{PC}=1^{-+}$) which has $\Pi_u$ excited glue. This
corresponds to  a width of around 100 MeV for the transition $H \to
\chi_b + f_0$, whereas when $M$ is an $\eta$ or $\eta'$ meson the
transition strength is found to be less than a few MeV.
 There will be modifications to this analysis coming from corrections to
the heavy quark limit  (of order $1/m_Q$ where $m_Q$ is the heavy quark
mass) which might allow hybrid meson transitions to  $B \bar{B}$, etc,
but these have not been evaluated yet.

 In this heavy quark (or static) limit,  the spin-exotic and non
spin-exotic hybrid  mesons are degenerate. For the latter, however,  the
interpretation of any observed states is less clear cut, since they 
could be conventional quark antiquark states. Moreover, the non
spin-exotic  hybrid mesons can mix directly (i.e. without emission of any
meson $M$) with conventional quark antiquark states once  one takes into
account corrections (of order $1/M_Q$) to the static approximation 
applicable for  heavy quarks with physical masses. 

 It is encouraging that the decay width comes out as relatively 
small, so that the  spin-exotic hybrid states should show up 
experimentally as sufficiently narrow resonances to be detectable.
 This decay analysis does not take into account heavy quark motion or
spin-flip  and these effects will be significantly more important for
charm quarks than for $b$-quarks.

\section{CONCLUSIONS}

  It is possible to extract information about hadronic decays from 
euclidian lattice studies. The most rigorous method is to  measure the
small energy shifts in the two body states precisely as the  lattice
volume is varied. This then allows the phase shift to be extracted 
which gives information on resonance properties. This energy shift  is
only enabled in dynamical fermion simulations if quark-antiquark
production is  needed in the decay. This makes this a very difficult 
route to follow in practice.

 We have also sketched an alternative which is less rigorous but  will
enable estimates to be obtained more readily. This direct method was 
tested in the case of $\rho \to \pi \pi$ where it gave consistent
results  and, moreover, results in agreement with experiment. This is a
good indication that  the method is reliable in practice. The method
works best when  the lattice transition amplitude ,$x$, is relatively
small and it is encouraging  that the method gives good indications when
$ax \approx 0.06$. 
 We then presented an application to hybrid meson decays: a case where
the  experimental result is not known, but where the magnitude of $x$ is
indeed small.  Our result is of experimental relevance:  the knowledge
of the magnitude of the decay width confirms that these hybrid
 states are likely to be accessible to straightforward study.

\section{ACKNOWLEDGEMENT}

 I have benefited from helpful discussions with my colleagues in the
UKQCD collaboration, especially Craig McNeile.

\end{document}